\documentclass{article}

\usepackage{arxiv}

\usepackage[utf8]{inputenc} 
\usepackage[T1]{fontenc}    
\usepackage{hyperref}       
\usepackage{url}            
\usepackage{booktabs}       
\usepackage{amsfonts}       
\usepackage{nicefrac}       
\usepackage{microtype}      
\usepackage{lipsum}		
\usepackage{graphicx}
\usepackage{natbib}
\usepackage{doi}
\usepackage{mathtools}

\title{Investigating certain choices of CNN configurations for brain lesion segmentation}

\author{ 
    \hspace{1mm}Masoomeh~Rahimpour 
    \thanks{Corresponding author: \texttt{masoomeh.rahimpour@kuleuven.be} } \\
	Department of Imaging and Pathology\\
	KU Leuven, Belgium \\
	\And
	\hspace{1mm}Ahmed~Radwan\\
	Department of Imaging and Pathology\\
	KU Leuven, Belgium \\
    \And
	\hspace{1mm}Henri~Vandermeulen\\
	Department of Radiology \\
	UZ Leuven, Belgium \\
	\And
	\hspace{1mm}Stefan~Sunaert \\
	Department of Imaging and Pathology\\
	KU Leuven, Belgium \\
	\And
    \hspace{1mm}Karolien~Goffin \\
	Nuclear Medicine\\
	UZ Leuven, Belgium \\
	\And
	\hspace{1mm}Michel~Koole \\
	Department of Imaging and Pathology\\
	KU Leuven, Belgium \\
}

\date{}

\renewcommand{\headeright}{Technical Report}
\renewcommand{\undertitle}{Technical Report}
\renewcommand{\shorttitle}{CNN configurations for brain lesion segmentation}

\hypersetup{
pdftitle={Investigating certain choices of CNN configurations for lesion segmentation tasks},
pdfauthor={Masoomeh.~Rahimpour, Karolien.~Goffin, Michel.~Koole},
pdfkeywords={},
}

\DeclareUnicodeCharacter{2212}{-}

\begin{document}
\maketitle

\begin{abstract}
Brain tumor imaging has been part of the clinical routine for many years to perform non-invasive detection and grading of tumors. Tumor segmentation is a crucial step for managing primary brain tumors because it allows a volumetric analysis to have a longitudinal follow-up of tumor growth or shrinkage to monitor disease progression and therapy response. In addition, it facilitates further quantitative analysis such as radiomics. Deep learning models, in particular CNNs, have been a methodology of choice in many applications of medical image analysis including brain tumor segmentation. In this study, we investigated the main design aspects of CNN models for the specific task of MRI-based brain tumor segmentation. Two commonly used CNN architectures (i.e. DeepMedic and U-Net) were used to evaluate the impact of the essential parameters such as learning rate, batch size, loss function, and optimizer. The performance of CNN models using different configurations was assessed with the BraTS 2018 dataset to determine the most performant model. Then, the generalization ability of the model was assessed using our in-house dataset. For all experiments, U-Net achieved a higher DSC compared to the DeepMedic. However, the difference was only statistically significant for whole tumor segmentation using $FLAIR$ sequence data and tumor core segmentation using $T_{1w}$ sequence data. Adam and SGD both with the initial learning rate set to $10^{−3}$ provided the highest segmentation DSC when training the CNN model using U-Net and DeepMedic architectures, respectively. No significant difference was observed when using different normalization approaches. In terms of loss functions, a weighted combination of soft Dice and cross-entropy loss with the weighting term set to $0.5$ resulted in an improved segmentation performance and training stability for both DeepMedic and U-Net models.
\end{abstract}


\section{Introduction}
Gliomas are the most common primary neoplasms in the brain~\citep{jin2020artificial}. Patients with high grade gliomas (HGG) which grow fast, infiltrate easily, and recur frequently, have a short survival rate while patients with low grade gliomas (LGG) which are slower growing lesions have a better prognosis. For both tumor types, extensive neuro-imaging protocols are being used to identify the tumor type, assess disease progression, optimize treatment plans such as surgical procedures and finally evaluate therapy response, including the detection of recurrences~\citep{alexiou2009glioma}. 
The precise delineations of the tumor can be obtained manually by expert readers. However, manual delineation based on high-resolution and multi-sequence 3D MRI data is a very time-consuming task, which also requires a high level of experience.

During recent years, considerable attention has been given to brain lesion segmentation resulting in different automatic segmentation methods. The superior predictive power of machine learning and more specifically deep learning models have made them the state-of-the-art approach for various aspects of brain image analysis, including, 1) tumor segmentation tasks to allow a further downstream analysis including radiomics or assessment of volumetric changes to evaluate disease progression, and 2) predictive tasks to identify the tumor type and predict treatment response, survival rate, or recurrences~\citep{jin2020artificial}.

Deep learning models, in particular CNNs, have been used to perform both of these tasks. Several model architectures along with different loss functions and optimization approaches have been developed to effectively train deep learning models such as VGG~\citep{simonyan2014very}, ResNet~\citep{he2016deep}, DenseNet~\citep{huang2017densely}, or Inception~\citep{szegedy2016rethinking}. Among these, U-Net~\citep{cciccek20163d} and DeepMedic~\citep{kamnitsas2017efficient} are the most commonly used model architectures for medical image segmentation. 

Choosing the optimal architecture and training configuration for a specific task is crucial to obtain maximum model performance and generalizability~\citep{isensee2021nnu}. Furthermore, high performance on standardized, publicly available data cannot always be translated to high clinical performance. This so-called domain shift or generalization is one of the main challenges of applying deep learning models in a clinical setting~\citep{guan2021domain}. 

In this study, we investigated several design choices to adapt DeepMedic and U-Net, for the specific task of MRI-based brain tumor segmentation. Different combinations of MRI sequences were used to train both models to perform the whole tumor and tumor core segmentation tasks. For whole tumor segmentation, $T_{1}$-weighted ($T_{1w}$), $T_{2}$-weighted ($T_{2w}$), fluid-attenuated inversion recovery ($FLAIR$) and any combination of these sequence data were used to train and test the models, while $T_{1w}$ or/and $T_{1}$-contrast-enhanced ($T_{1ce}$) sequence data were used for tumor core segmentation. A high-resolution $T_{1w}$ sequence is the most frequently used MRI sequence in the clinical setting. On the other hand, $T_{2w}$ and $FLAIR$ sequence data provide more relevant information for the segmentation of whole tumor~\citep{menze2014multimodal}, while $T_{1ce}$ sequence provides much more specific information to differentiate the viable tumor tissue from the healthy tissue~\citep{upadhyay2011conventional}. Essential parameters such as learning rate, batch size, class sampling strategy, loss function, and optimizer were evaluated for both model architectures and compared to determine the most performant model. Then, the generalization ability of the model was assessed by using an additional in-house dataset. Because of the limited size of the in-house dataset, transfer learning was considered to improve segmentation performance by starting from pre-trained models using the BraTS 2018 dataset.

\section{Methods and materials}\label{sec:{Methods}}
In this section, we first describe the datasets that were used to evaluate the performance of DeepMedic and U-Net for MRI-based brain tumor segmentation. Next, we discuss the two CNN models that were used for our experiments on brain tumor segmentation. Finally, we give an overview of the experiments which were performed to explore the impact of different parameters on the performance of the two CNN models.

\subsection{Data}
\textbf{BraTS 2018}

The BraTS dataset consists of MRI scans of patients with brain tumors collected from multiple imaging centers with different MRI scanners and using different imaging protocols. This resulted in a vastly heterogeneous dataset of 285 multi-contrast and multi-sequence MRI scans in total, consisting of 210 HGG cases and 75 LGG cases~\citep{menze2014multimodal}. Each case contained four 3D MRI sequences including $T_{1w}$, $T_{1ce}$, $T_{2w}$ and $FLAIR$ data, all of them rigidly aligned with the common anatomical SRI24 brain template~\citep{rohlfing2010sri24}. All MRI data have been skull stripped, and resampled to 1 mm isotropic resolution, resulting in an image size of $240\times 240\times 155$. The manual delineation was performed by at least two experts with a maximum of four experts, and the results were fused to obtain a consensus segmentation for each lesion in four tumor tissue classes: edema, the tumor core which corresponds to the enhanced part of the tumor on $T_{1ce}$, the part of the tumor which is not enhanced on $T_{1ce}$ and the necrotic tumor core.

\textbf{In-house data}

This dataset consisted of pre-operative, multi-sequence MRI data of 37 patients with gliomas. These data were acquired at UZ Gasthuisberg (Leuven, Belgium) as part of a study to evaluate a virtual brain grafting (VGB) pipeline for improving brain parcellation in the presence of large lesions~\citep{radwan2021virtual}. For each patient, four 3D MRI sequences were acquired ($T_{1w}$, $T_{1ce}$, $T_{2w}$, $FLAIR$), spatially normalized with the SRI24 brain template, and co-registered with the $T_{1w}$ sequence data while data were resampled to a 1 mm isotropic resolution. The study was approved by the local research ethics committee at UZ/KU Leuven. Ground truth annotations were generated by two neuro-radiologists, using ITK-SNAP (v3.8.0)~\citep{yushkevich2006user} and the protocol described in~\citep{yushkevich2019user} for multi-sequence semi-automatic lesion segmentation. Radiologist 1 has 6 years of experience in brain imaging while radiologist 2 has 4 years of experience.

\textbf{Data preprocessing}

All input images from both datasets were cropped using a fixed size bounding box ($143\times 196\times 144$) to ensure that voxels belonging to the brain mask were included in the image. The bounding box was defined based on the SRI24 brain template that was used to spatially normalize the BraTS 2018 data. 

\subsection{CNN models}

\textbf{DeepMedic}

\begin{figure*}[h]
\centering
\footnotesize
\includegraphics[width=1.0\textwidth]{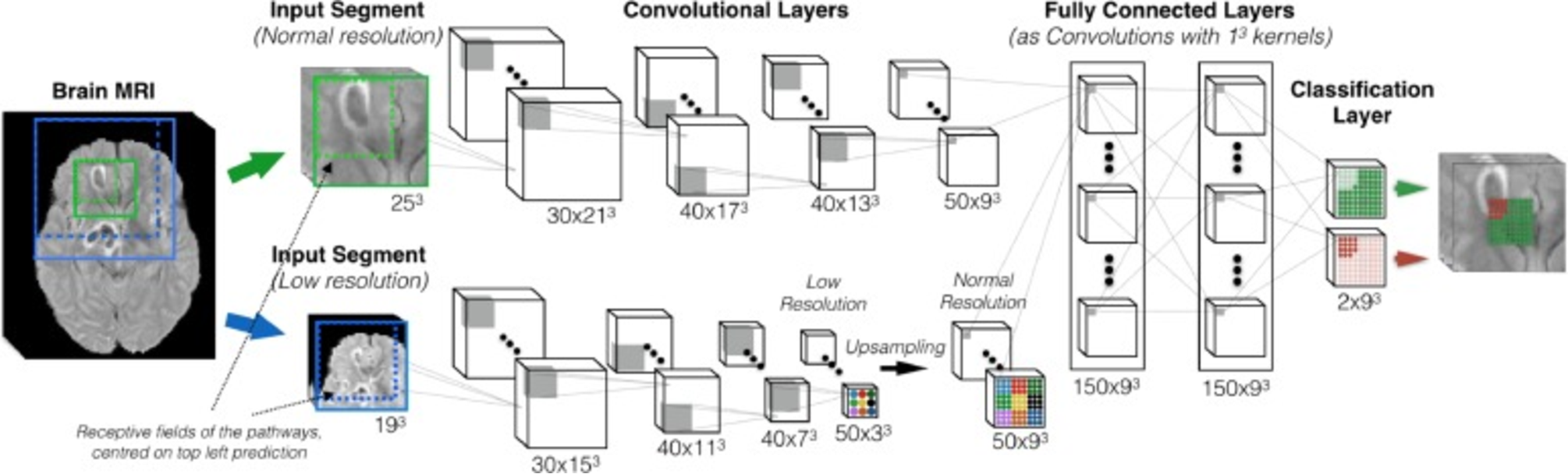}\\ 
\caption[Schematic overview of DeepMedic architecture]{Schematic overview of DeepMedic architecture with two convolutional pathways proposed for brain lesion segmentation~\citep{kamnitsas2017efficient}.} \label{fig:DM}
\end{figure*}

DeepMedic is a dual pathway 3D CNN designed for lesion segmentation~\citep{kamnitsas2017efficient}. It consists of two parallel convolutional pathways at different resolutions. Adding a low-resolution pathway that operates on down-sampled input images allows the model to achieve a large receptive field and include more contextual information for the final classification while the local information is preserved in the high-resolution pathway. By using input images with a larger size than the model’s receptive field, the model generates multiple predictions for each voxel. Compared to other patch-based approaches which use overlapped patches to obtain reliable predictions for each voxel, this strategy significantly reduces the computational cost. The model consists of 11 layers including 8 convolutional layers, 2 fully-connected layers, and 1 classification layer as depicted in Figure~\ref{fig:DM}. The convolutional layer performs a convolution with a $3\times 3\times 3$ kernel followed by batch normalization and a rectified linear unit (ReLU) activation function. The fully-connected layers which are used to concatenate the different pathways and map the feature maps to the classification layer are implemented as convolution layers with a $1\times 1\times 1$ kernel. To prevent an internal covariate shift and to speed up the training, batch normalization is applied to normalize the feature map activations at every optimization step. Training is regularized by using dropout~\citep{srivastava2014dropout} with a rate of 2\% for all convolutional layers, in addition to a 50\% dropout rate for the two fully connected layers.

DeepMedic was trained using 3D patches with the centroids of the patches sampled uniformly inside the brain mask and with the second pathway using images that were downsampled three times. The patches were randomly sampled from the 3D images with a size of $25\times 25\times 25$. In each iteration, the gradient was calculated using a random batch of 60 patches selected from a maximum of 10 subjects. 

\textbf{U-Net}

\begin{figure*}[h]
\centering
\footnotesize
\includegraphics[width=1.0\textwidth]{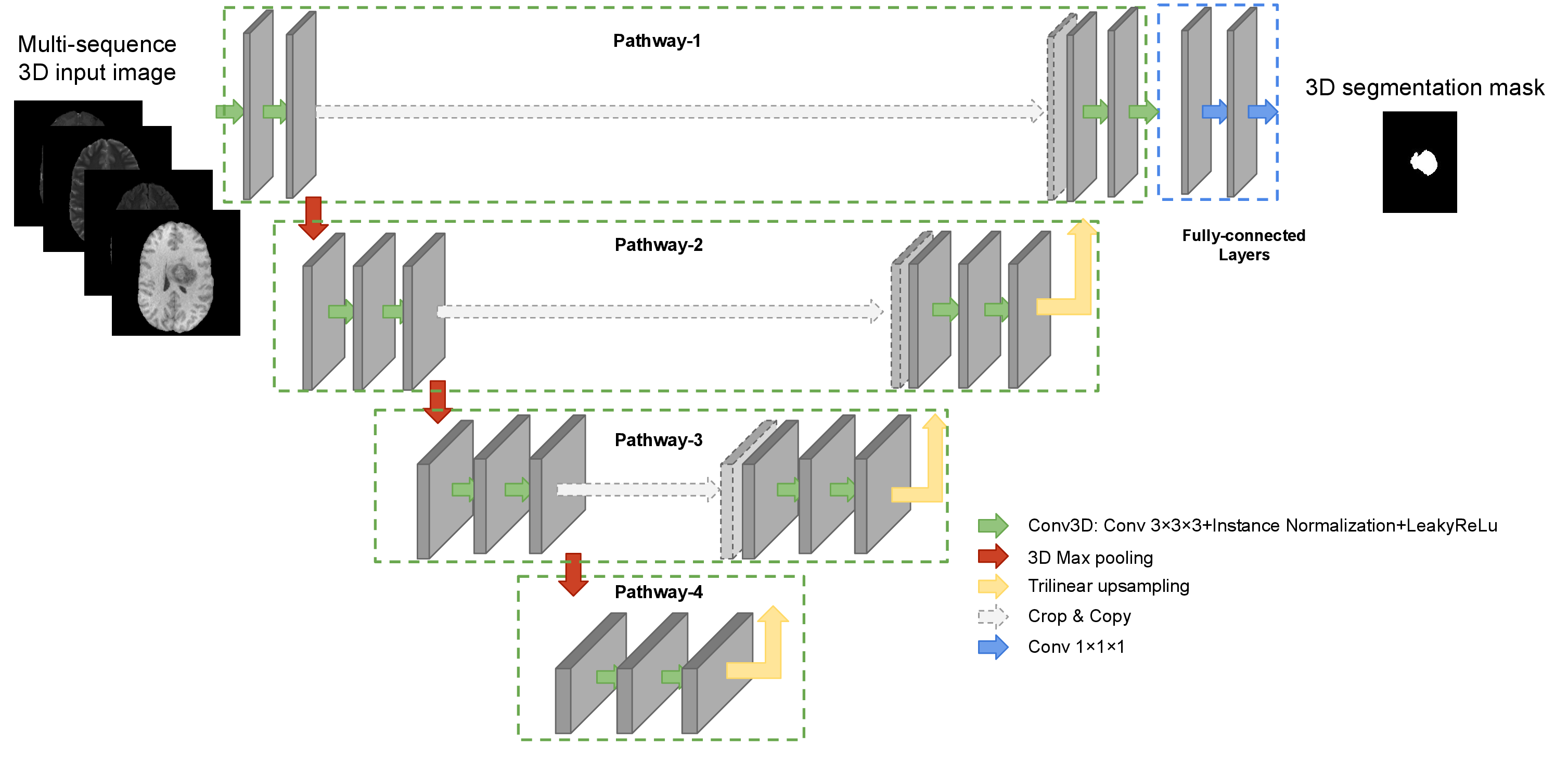}\\ 
\caption[Schematic overview of U-Net architecture]{Schematic overview of U-Net architecture with four resolution pathways and three skip connections.} \label{fig:UNET}
\end{figure*}

U-Net has an encoder-decoder architecture which was originally developed for biomedical image segmentation~\citep{Cicek2016}. The encoder branch is constructed by stacking four pathways, each consisting of two convolutional layers with a $3\times 3\times 3$ kernel. All convolutional layers are followed by an instance normalization and a ReLU activation function. After each convolutional block, a max-pooling layer with a $3\times 3\times 3$ kernel is applied to reduce the size of feature maps by the factor of three, while the number of kernels in convolutional layers is doubled each time. The decoder branch contains the same convolutional blocks each followed by an up-sampling layer to reconstruct the original image dimension. The number of feature maps after each up-sampling layer is halved. In each pathway, the feature maps of the encoder and decoder branches are concatenated via the skip connections. Two fully-connected layers followed by a softmax layer are added as final layers to classify the image voxels into healthy or tumoral tissue. Similar to the DeepMedic architecture, the fully-connected layers are implemented as the 3D convolutions with a $1\times 1\times 1$ kernel. Same padding was used in all convolutional layers. Training was regularized by using dropout with a 50\% rate for the final two fully connected layers. A schematic overview of U-Net architecture is shown in Figure~\ref{fig:UNET}.

U-Net was trained using the full images with the size of $143\times 196\times 144$ and the batch size of 2 where the output size of predictions was same as the size of the input images and the loss function was calculated over all image voxels.

\textbf{Training}

The input to both models was a 5-dimensional tensor which was obtained by the concatenation of a specific number of batches with each batch consisting of single or multiple 3D MRI sequences.
Both models were initialized by sampling a normal distribution, centered around zero with standard deviation $\sqrt{2/n^{in}_{l}} (N(0, \sqrt{2/n^{in}_{l}}))$ with $n^{in}_{l} = C_{l-1}\prod_{i=\{x,y,z\}} k_{l}^{(i)}$, the number of weights connecting each neuron of layer $l$ to its input, $C_{l}$ the number of feature maps in layer $l$ and $k_{l}^{\{x,y,z\}}$ the size of the kernel in the respective spatial dimension~\citep{he2015delving}. 

During training, a combination of random scaling, in-plane rotation, translation, and lateral flipping was used for on-the-fly data augmentation. Images were reflected along the axial, coronal, and sagittal axes with a 50\% probability. Furthermore, a random intensity shift and scale, sampled from a Gaussian distribution $N(0,0.05)$ and $N(0,0.01)$, respectively were applied to the training data for all MRI sequences. 

We used the hold-out method to split the data into 80\% training and 20\% test data. During training, we additionally performed five-fold cross-validation to optimize the number of epochs and different hyperparameters. For each fold, the model with the highest validation DSC was chosen. The final prediction for the test dataset was generated by averaging the predictions obtained by the five models resulting from the five training folds. The same training and test dataset distribution were used to train and test the DeepMedic and U-Net models. 

All the experiments using both CNN architectures were implemented in Tensorflow (v1.9.0). All computation was performed on the Flemish supercomputer (CentOS Linux 7) using 2 NVIDIA P100 GPUs (CUDA v11.0, GPU driver v450.57) and 1 Intel Skylake CPU (18 cores).

\subsection{Model optimization}
We evaluated different aspects to further optimize the performance of the two CNN models. First, we evaluated different optimization schemes for training the CNN model. Next, we considered different approaches for intensity normalization of the MRI input data. Finally, we considered different loss functions and evaluated the impact on the model performance. For each of these aspects, the performance of the trained model was evaluated using the BraTS 2018 dataset by comparing the DSC between the predicted and ground truth segmentations of the validation dataset.

\textbf{Optimization schemes}

For training, both stochastic gradient descent (SGD) with $0.99$ Nesterov momentum and Adam optimizer~\citep{kingma2014adam} were considered using default Keras setting (v 2.2.4) with different initial learning rates ($10^{-4}$, $10^{-3}$, $10^{-2}$). The learning rate was reduced by a factor of 5 when the validation loss for the full images did not improve anymore and training was stopped when the validation loss increased. 

\textbf{Intensity normalization}

The intensity values of each input image were normalized to zero mean and unit standard deviation while considering only non-zero voxels to calculate the mean and standard deviation. In addition, the maximum-minimum, and 95-th percentile of the intensity values per image were used for intensity normalization. Furthermore, the effect of bias field correction using the N4 algorithm~\citep{tustison2010n4itk} implemented by the advanced normalization tools (ANTs) has been investigated to correct intensity inhomogeneity. 

\textbf{Loss functions}

With the lesions occupying a considerably smaller volume relative to the background (i.e. healthy tissue), we faced the issue of class-imbalance when performing the tumor segmentation task using DeepMedic. To deal with this issue, a weighted sampling approach was performed such that a certain portion of batches consisted of the patches centered on healthy tissue and the remaining located on the tumoral tissues. Furthermore, the effect of using different loss functions to address this issue was investigated. Three groups of loss functions including the cross-entropy-based loss ($L_{CE}$) (i.e. weighted cross-entropy ($L_{WCE}$) and focal loss ($L_{Focal}$))~\citep{lin2017focal}, metric-sensitive loss (i.e. soft Dice loss ($L_{SD}$)) and the weighted combination of $L_{CE}$ and $L_{SD}$ ($L_{Comb}$) were used to train the CNN models. The loss functions are defined as follows:

\begin{equation}
  L_{WCE} = -\frac{1}{N} \Sigma_{n=1}^{N} \omega\log(p_{n}) + (1-\omega)\log(1-p_{n}),
\end{equation}

where $p_{n}$ and $(1-p_{n})$ represent the predicted probability of the foreground and background classes, respectively and $\omega$ is the weighting term attributed to the foreground class. This function can be written in a short form:

\begin{equation}
  L_{WCE}(p_{t}) = -\omega_{t} \log(p_{t}), 
\end{equation}

\begin{equation}
    p_{t} = 
    \begin{cases}
    p_{n}, & if~~ y=1 \\
    1-p_{n}, &   otherwise.\\
    \end{cases}
\end{equation}

Given this notation, $L_{Focal}$ is defined as follows: 

\begin{equation}
    L_{Focal}= -\omega_{t} (1-p_{t})^{\gamma} \log(p_{t}),
\end{equation}

with $\gamma$, the focusing parameter which reduces the relative loss for the easy, well-classified examples and puts more focus on hard misclassified examples.

\begin{equation}
    L_{SD} = 1 - \frac{\Sigma_{i} \hat{y_{i}}y_{i} + \epsilon}{\Sigma_{i} \hat{y_{i}} + \Sigma_{i} y_{i} + \epsilon},
\end{equation}

where $\hat{y_{i}}$ and $y_{i}$ represent the predicted and the ground truth label for voxel $i$, respectively, and $\epsilon$ is the smoothing term to avoid division by zero if $\Sigma_{i} \hat{y_{i}}$ and $\Sigma_{i} y_{i}$ are empty. 

Finally, $L_{Comb}$ is defined as:
\begin{equation}
   L_{Comb} = \alpha L_{SD} + (1-\alpha)L_{CE}, 
\end{equation}

where $\alpha$ is a parameter to balance the weight of loss terms.

A grid search with the range of [0.25, 0.5, 0.75, 1] was used to choose $\omega$ when using $L_{WCE}$. When using $L_{Focal}$, a grid search of [0.25, 0.5, 0.75, 1] and [1, 1.5, 2, 2.5] was used to choose $\omega$ and $\gamma$, respectively. When using $L_{Comb}$, the weighting term, $\alpha$, was selected with a grid search with the range of [0.1:0.9] with the steps of 0.1. The effect of using different loss functions to train the model was assessed for the task of whole tumor segmentation on the BraTS 2018 test dataset when using all four MRI sequence data to train the model.

\subsection{Robustness analysis}
To analyze the robustness and stability of both models, two sets of experiments were performed by setting a varying number of model parameters and random initialization to train the CNN models. Hypothetically, larger models have superior performance and generalizability~\citep{ba2014deep}. Nevertheless, limited computational resources and memory constraints of the current hardware limit the model size. We trained four sets of models by varying numbers of model parameters by using a different number of kernels in convolutional and fully connected layers for each CNN architecture. Next, both CNN models were retrained 10 times within a 5-fold cross-validation experiment while being randomly initialized and segmentation performance was evaluated by determining the DSC (\%) on the validation datasets. This robustness analysis was performed on the BraTS 2018 dataset for whole tumor segmentation when using four MRI sequence data as the input to train the models.

\subsection{U-Net vs DeepMedic}
Using the optimal configuration, an extensive analysis was performed on multi-sequence brain tumor segmentation using both BraTS 2018 and our in-house data. For the evaluation, both whole tumor and tumor core segmentation were considered as segmentation tasks. 

A Wilcoxon signed-rank test was used to compare DSC obtained by DeepMedic and U-Net and to determine whether there was a statistically significant difference in segmentation performance between the two CNN models.

\subsection{Generalizability of U-Net}
An in-house dataset was used to evaluate the generalizability of U-Net by considering the model configuration which achieved the highest DSC. We first considered the U-Net model trained with BraTS 2018 dataset and used this pre-trained model as such to evaluate the performance of the different models on the in-house test datasets. Next, we considered transfer learning to fine-tune the pre-trained models. For this purpose, a pre-trained model was loaded after which all layers were fine-tuned on the new data domain using 18 randomly selected in-house datasets. Meanwhile, the remaining 19 in-house datasets were used for testing. A Wilcoxon signed-rank test was used to compare DSC obtained before and after transfer learning to determine whether transfer learning resulted in statistically significant improvement of the segmentation performance.

\section{Results}
\subsection{Model optimization}
No statistically significant difference (p-value=0.4) was observed when performing the intensity normalization using zero mean and unit standard deviation, maximum-minimum, and 95-th percentile of the intensity values per image. Nevertheless, when applying bias field correction, the segmentation performance was degraded based on the validation dataset. Therefore, we did not consider this preprocessing step to train the final model. For the final evaluations, the 95-th percentile of the intensity values per image was used for intensity normalization.
Regarding the optimizer and learning rate, Adam with the initial learning rate set to $10^{-3}$ provided the highest segmentation DSC to train the U-Net model. When training the DeepMedic model, SGD with the initial learning rate set to $10^{−3}$ was considered the most appropriate optimizer. 

Table~\ref{tab:DSCLF} represents the average DSC with the weighting term (i.e. $\alpha$, $\omega$, and $\gamma$) corresponding to the highest DSC. The highest DSC for both DeepMedic and U-Net models was achieved by using $L_{Comb}$ and setting the weighting term to $0.5$. For all experiments, the performance of U-Net was superior to DeepMedic.

\begin{table*}[h!]
\setlength{\tabcolsep}{4pt} 
\scriptsize
\caption[Average DSC for whole tumor segmentation when using different loss functions]{Average DSC (\%) for whole tumor segmentation when using different loss functions to train the CNN models (i.e. DeepMedic and U-Net). Four MRI sequence data from the BraTS 2018 dataset were used to train the CNN models.}\label{tab:DSCLF}
\begin{tabular*}{\linewidth}{@{\extracolsep{\fill}} cccc}
\toprule
Loss function & Weighting term(s) & DeepMedic & U-Net \\
\midrule
$L_{WCE}$ & $\omega$ = 0.75 & 84.86 & 86.53 \\
$L_{Focal}$ & $\omega$= 0.75, $\gamma$=1.5 & 85.28 & 84.34 \\
$L_{SD}$ & $-$ &	87.02 & 89.08 \\
$L_{Comb}$ &  $\alpha$ = 0.5 & 	87.33 & 89.97 \\
\bottomrule
\end{tabular*}
\end{table*}

\subsection{Robustness analysis}
In terms of model size, the number of kernels were chosen as [30, 30, 40, 40, 40, 40, 50, 50] for the convolutional layers and [150, 150] for the fully-connected layers when using the DeepMedic architecture resulting in 663,491 trainable parameters. When reducing the model size to 50\% and 25\% of this size, the segmentation DSC significantly degraded, while by increasing the number of trainable parameters by 50\%, using the same patch size no statistically significant improvement was observed in terms of segmentation performance. 
U-Net was implemented by starting with 20 and 40 kernels, both with a kernel size of $3\times 3\times 3$, for the convolutions in the first pathway and with 20 kernels for the fully connected layers resulting in 3,878,841 trainable parameters. When reducing the model size by 50\% and 25\%, the DSC significantly decreased. Since we used full images as input to train the model, increasing the number of trainable parameters was not feasible with the available GPU resources because of the memory constraints.
When re-training the model starting from a random initialization, the standard deviation of the average DSC over the 10 training sessions was 0.49\% and 0.36\% for DeepMedic and U-Net, respectively. 

\subsection{U-Net vs DeepMedic}
Table~\ref{tab:UNETvsDM} lists the DSC obtained by DeepMedic and U-Net for whole tumor and tumor core segmentation using the BraTS 2018 test dataset. The results are reported when using different sets of MRI sequences to train and test the models for both tasks. For all experiments, U-Net achieved a higher DSC compared to the DeepMedic. The DSC significantly improved using U-Net for whole tumor segmentation using $FLAIR$ sequence data and tumor core segmentation using $T_{1w}$ sequence data. 

\begin{table*}[h!]
\setlength{\tabcolsep}{4pt} 
\scriptsize
\caption[Segmentation DSC on BraTS 2018 test datasets when using DeepMedic and U-Net]{Average$\pm$standard deviation of DSC (\%) for whole tumor and tumor core segmentation on BraTS 2018 test datasets when using DeepMedic and U-Net architectures when different MRI sequence data were used to train the models. The bold font denotes statistically significant improvement according to the Wilcoxon test (p < 0.05).}\label{tab:UNETvsDM}
\begin{tabular*}{\linewidth}{@{\extracolsep{\fill}} cccc}
\toprule
Segmentation task &	MRI sequence &	DeepMedic &	U-Net \\
\midrule
& $T_{1w}$ & 68.73$\pm$13.17 & 71.37$\pm$15.08 \\
& $T_{2w}$ & 77.42$\pm$12.29 & 80.73$\pm$.01 \\
& $FLAIR$ & 78.47$\pm$13.52	& \textbf{82.05$\pm$5.99}\\
Whole tumor	& $T_{1w}$ +$T_{1w}$ & 80.07$\pm$12.09 & 81.04$\pm$11.38 \\
& $T_{1w}$+$FLAIR$ & 81.48$\pm$7.73	& 84.62$\pm$10.35 \\
& $T_{2w}$+$FLAIR$ & 82.66$\pm$9.08 & 84.69$\pm$5.98 \\
& $T_{1w}$+$T_{2w}$+$FLAIR$	& 84.48$\pm$7.73 & 85.63$\pm$6.31 \\
\midrule
& $T_{1w}$ & 58.62$\pm$19.52 & \textbf{64.80$\pm$18.46} \\
Tumor core & $T_{1ce}$ & 69.12$\pm$16.08 & 70.1$\pm$16.88 \\
& $T_{1w}$+$T_{1ce}$ & 78.89$\pm$13.96 & 81.10$\pm$13.07 \\
\bottomrule
\end{tabular*}
\end{table*}

\subsection{Generalizability of U-Net}
Generalizability was only assessed for U-Net. Table~\ref{tab:UNETGEN} lists the average DSC obtained by U-Net for whole tumor and tumor core segmentation using the in-house dataset without and with transfer learning. The same MRI sequence data that were used for the DeepMedic vs U-Net evaluation using the BraTS 2018 dataset, were also used for the evaluation using the in-house dataset. DSC was determined by using the manual delineations by two radiologists (R1 and R2) as the ground truth labels. The inter-radiologist agreement was 85.38$\pm$16.76\% and 77.10$\pm$13.32\% for whole tumor and tumor core segmentation, respectively. This was closely approximated by the best performant U-Net model when using the combination of $T_{1w}$, $T_{2w}$, $FLAIR$ sequence data for whole tumor segmentation and combination of $T_{1w}$, $T_{1ce}$ sequence data for tumor core segmentation when applying the transfer learning.

When applying transfer learning, the segmentation performance was significantly improved for all experiments. Nevertheless, when using the manual delineations of the two radiologists as the ground truth labels, no statistically significant difference was observed in the model performance for whole tumor and tumor core segmentation without or with transfer learning.

\begin{table*}[h!]
\setlength{\tabcolsep}{4pt} 
\scriptsize
\caption[Segmentation DSC with and without transfer learning on the in-house test datasets when using U-Net model]{Average$\pm$standard deviation of DSC (\%) for whole tumor and tumor core segmentation with and without transfer learning on the in-house test datasets when using U-Net model. Different MRI sequence data were used to train the U-Net model and the results were reported when using the manual delineations obtained by the two radiologists (R1 and R2) as the ground truth labels.}\label{tab:UNETGEN}
\begin{tabular*}{\linewidth}{@{\extracolsep{\fill}} cccc}

a. No transfer learning \\
\toprule
Segmentation task & MRI sequence & \multicolumn{2}{c}{DSC (\%)}\\
\midrule
& & R1 & R2 \\
\midrule
& $T_{1w}$ & 65.33$\pm$12.49 & 64.17$\pm$12.56  \\
& $T_{2w}$ & 75.78$\pm$7.72	& 73.59$\pm$28.25  \\
& $FLAIR$ & 79.72$\pm$7.55 & 78.48$\pm$14.21 8 \\
Whole tumor	& $T_{1w}$+$T_{2w}$	& 67.24$\pm$21.90 & 68.24$\pm$33.13  \\
& $T_{1w}$+$FLAIR$ & 79.02$\pm$8.28 & 79.81$\pm$15.24  \\
& $T_{2w}$+$FLAIR$	& 80.09$\pm$5.97 & 80.57$\pm$6.89  \\
& $T_{1w}$+$T_{2w}$+$FLAIR$	& 81.88$\pm$16.7 & 81.6$\pm$18.66  \\ 
\midrule
& $T_{1w}$	& 58.85$\pm$21.71 & 58.72$\pm$24.28 \\
Tumor core  & $T_{1ce}$	& 62.35$\pm$21.58 & 61.45$\pm$21.45 \\
& $T_{1w}$+$T_{1ce}$ & 62.74$\pm$21.79 & 60.37$\pm$20.17 \\
\bottomrule
\end{tabular*}
  
\begin{tabular*}{\linewidth}{@{\extracolsep{\fill}} cccc}
\\
b. Transfer learning \\
\toprule
Segmentation task & MRI sequence & \multicolumn{2}{c}{DSC (\%)}\\
\midrule
& & R1 & R2 \\
\midrule
& $T_{1w}$ &  73.60$\pm$7.65 & 72.13$\pm$13.66 \\
& $T_{2w}$ &  80.12$\pm$4.33 & 76.69$\pm$8.49 \\
& $FLAIR$  &  82.30$\pm$15.64 & 82.74$\pm$18 \\
Whole tumor	& $T_{1w}$+$T_{2w}$	&  74.56$\pm$22.33 & 73.38$\pm$25.71 \\
& $T_{1w}$+$FLAIR$ & 82.56$\pm$8.28 & 83.63$\pm$16.35 \\
& $T_{2w}$+$FLAIR$ & 84.46$\pm$4.12 & 84.88$\pm$5.82 \\
& $T_{1w}$+$T_{2w}$+$FLAIR$	& 85.04$\pm$5.22 &	85.46$\pm$9.14 \\ 
\midrule
& $T_{1w}$ &  64.35$\pm$13.20 & 63.33$\pm$14.23 \\
Tumor core  & $T_{1ce}$	& 71.15$\pm$12.32 &	70.84$\pm$12.95 \\
& $T_{1w}$+$T_{1ce}$ & 71.90$\pm$14.24 & 71.46$\pm$15.12 \\
\bottomrule
\end{tabular*}
\end{table*}

\section{Discussion and conclusion}
In this study, we explored the impact of different model configurations on the performance of the CNN model. This investigation is specifically important because the design of a CNN model directly affects the model performance and therefore the uncertainty of the predictions. For instance, the memorization capacity of the model is determined based on the model size (i.e. the number of trainable parameters). Therefore, using a suboptimal model size could lead to under- or over-fitting. In addition, access to computation resources is another limiting factor to define the model size.  We used DeepMedic and U-Net as the two commonly used CNN architectures to segment tumor sub-regions using multi-sequence brain MRI data. our results were in line with the global challenge scores from more recent and successful methods such as nn-UNet although these newer methods were developed using the much larger BraTS 2021 data with improved labels. To assess the generalizability of a trained U-Net model, we used an in-house dataset in addition to the publicly available BraTS 2018 dataset. Ground truth labels of our in-house dataset were delineated by two independent radiologists such that the inter-radiologist agreement could also be determined and used for the performance comparison of the CNN models. We used one single set of ground truth labels to train the models. However, using the manual delineations by multiple radiologists as independent ground truth labels to train the model could also be assessed. This can be implemented with an encoder-decoder architecture where a shared encoder is used to extract the features from the input images while radiologist-specific decoders take the manual delineation of each radiologist as the ground truth labels to perform the classification. It allows the model to learn the intra-radiologist variability of ground truth labels resulting in feature maps that are less dependent on the radiologist.  

While there are multiple choices to set an optimal configuration of a CNN model, limiting these choices to a few categories allows for an optimized design and results in a robust performance~\citep{isensee2021nnu}. Model architecture, preprocessing, and training scheme are the main configuration aspects that we investigated. 

\textbf{The importance of model architecture}: All the configurations described earlier were explored for both DeepMedic and U-Net architectures. Using the U-Net model we always obtained a higher DSC, but the difference was only statistically significant for whole tumor segmentation with $FLAIR$ sequence data and tumor core segmentation using $T_{1w}$ sequence data. It can be hypothesized that applying a pooling operation on the feature maps within the encoder branch of the U-Net enables the model to extract the features at multiple resolutions, which could result in better capturing the contextual information, while in DeepMedic downsampling only happens once when using the input imaging in the low-resolution pathway. Nevertheless, the advantage of using the full image to train the model and the larger number of trainable parameters when using the U-Net architecture  cannot be neglected. 

According to the results we obtained, we chose the U-Net architecture with the model configurations resulting in the highest DSC as a starting point to train the CNN models used in our next studies. However, selecting a single optimal configuration to train a CNN model to perform a certain task is challenging, and a task-specific configuration is needed depending on the dataset proprieties and hardware specifications. 

\textbf{The importance of intensity normalization}: No substantial difference was observed in segmentation accuracy when applying different intensity normalization methods using the zero mean and unit standard deviation, maximum-minimum, and 95-th percentile of intensity values per image. 

\textbf{The importance of sampling and loss function}: When performing the patch-based training using DeepMedic, weighting sampling to enable the model to equally use the patches sampled from the foreground and background tissue resulted in the highest DSC compared to the scenario we did not apply the weighting sampling. In terms of loss functions, a weighted combination of soft Dice and cross-entropy loss with the weighting term of 0.5 resulted in an improved segmentation performance and training stability for both DeepMedic and U-Net models. However, we believe the weighting terms could vary for different applications. 

\textbf{The importance of patch size, batch size, and voxel spacing}: For the patch-based training (i.e. DeepMedic), the larger the patch size, the more contextual information is provided for the model. Furthermore, when defining the depth of the model it is crucial to set the receptive field size at least as large as the input patch size such that the contextual information is retained. Image resolution and batch size are mainly defined based on the availability of GPU memory. The larger the batch size, the more accurate gradient estimate is achieved. Given the memory constraints, when using the full images as the input to train the U-Net model we limited the batch size to 2 and the training appeared to be quite robust, while the batch size was set to 60 for the patch-based training using DeepMedic. We also observed that cropping the data to their non-zero regions improved the computational efficiency. In addition, training the segmentation model on downsampled input images increased the size of patches and enables the model to accumulate more contextual information. Although it could result in a small error, especially for the small segmentation targets, we did not observe a statistically significant difference in the segmentation performance when downsampling the input image to 2 mm voxel size~\citep{rahimpour2021improving}~\citep{rahimpour2022clinical}. Therefore, downsampled input images could be used to train larger CNN models such as the cross-modal knowledge/feature distillation (KD/FD) models~\citep{rahimpour2021cross} and the feature-level fusion (FLF) model~\citep{rahimpour2022visual}. Nevertheless, with unlimited GPU memory, it is recommended to train the models at full resolution with a patch size that covers the full image. 

\textbf{The importance of transfer learning}: We investigated the generalizability of the U-Net model using our in-house dataset. Due to the limited size of the in-house dataset, transfer learning was applied to fine-tune the CNN models pre-trained on BraTS 2018 data. When performing the transfer learning on a small set of new training data, the segmentation performance significantly improved compared to training the model from the scratch on a small set of new training data. Indeed, transfer learning improved the segmentation DSC up to 8\% and 11\% for whole tumor and tumor core segmentation, respectively with generally lower DSC variability and thus more consistent segmentation accuracy was achieved. 

\textbf{Other aspects}: Choosing the right metric to assess the performance of a segmentation model highly depends on the task for which the output of the segmentation model is used. Overlap ratio measures such as Jaccard, and Dice are the most common metrics to evaluate the accuracy of segmentation model~\citep{isensee2021nnu}. Nevertheless, if the segmentation output (i.e. predicted lesion) should be monitored for changes in size, mean or absolute volume difference is the metric to be calculated. On the other hand, if the segmentation output is used for treatment planning (i.e. radiation therapy), the shape fidelity of the segmentation output to the ground truth label is the most important metric for evaluation. In this case, the Hausdorff metric is useful where the boundary similarity between the segmentation output and ground truth label is measured. For all the experiments in the study, we used DSC as the evaluation metrics.

We did not apply any post-processing on the prediction segmentation maps. However, connected component analysis (CCA) or conditional random field (CRF) are among the post-processing approaches that were used in the literature to remove the false positive detections~\citep{kamnitsas2017efficient}.

\section*{Acknowledgment}
The authors would like to acknowledge the funding for the research presented in this study, provided by the European Union’s Horizon 2020 Research and Innovation Programme under the Marie Sklodowska-Curie grant agreement No 764458 (HYBRID project). The resources and services used in this work were provided by the VSC (Flemish Supercomputer Center), funded by the Research Foundation – Flanders (FWO) and the Flemish Government. Authors would also like to thank NVIDIA Corporation for donating a Titan X GPU.

\bibliographystyle{unsrtnat}
\bibliography{references} 

\end{document}